\documentstyle[12pt]{article}
\begin{document}

\begin{center}
{\Large\bf Non abelian bosonisation in three dimensional field theory}
\vskip 0.5in 
N. Banerjee\footnote{e-mail:narayan@bose.ernet.in}\\[.5cm]
 S.N.Bose National Centre for Basic Sciences\\
 Block JD, Sector III, Salt Lake City\\
Calcutta 700091, India\\[.5cm]
 R. Banerjee\footnote{On sabbatical leave from S.N. Bose National Centre
for Basic Sciences,Block JD, Sector III, Salt Lake , Calcutta -
700091, India. E-mail:rabin@if.ufrj.br}\\[.5cm]
 Instituto de Fisica\\
 Universidade Federal do Rio de Janeiro\\
 RJ 21945-970 Caixa Postal 68528, Brasil\\
 and\\
 S. Ghosh\\[.5cm]
 Gobardanga Hindu College\\
24-Parganas (North) West Bengal, India\\
\end{center}

\begin{abstract}
We develop a method based on the generalised St\"uckelberg prescription
for discussing bosonisation in the low energy regime of the SU(2)
massive Thirring model in 2+1 dimensions. For arbitrary
values of the coupling parameter the bosonised theory is
found to be a nonabelian gauge theory whose physical sector is
explicitly obtained. In the case of
 vanishing coupling this gauge theory
can be identified with the SU(2) Yang-Mills Chern-Simons theory in the
limit when the Yang-Mills term vanishes. Bosonisation identities for
the fermionic current are derived.
\end{abstract}
\newpage
The conventional ideas \cite{r1} of bosonising a (1+1) dimensional
fermionic theory have been recently extended in several directions.
These include a wide variety of topics like `smooth bosonisation'
\cite{r2,r3,r4},  connection between dual
transformations and bosonisation \cite{r5,r6}, bosonisation
in higher dimensions \cite{r6,r7,r8,r9,r10,r11} etc. Most of these
issues, however, are analysed in the context of abelian models. Thus,
inspite of this recent plethora of papers, not much insight has been
gained concerning nonabelian bosonisation, particularly in dimensions
greater than (1+1). There is a paper \cite{r10} which discussed the
bosonisation of the three-dimensional nonabelian massive Thirring
model (in the low energy regime) but definitive results were obtained
only in the weak coupling limit in which case the original model
reduced to a theory of free massive fermions.

In this paper we develop a general formalism of discussing
bosonisation in the low energy sector of nonabelian fermionic models
in higher dimensions. To pinpoint our analysis we consider the
bosonisation of SU(2) massive Thirring model in 2+1 dimensions. At the
end it will be evident that considering a larger group of
transformations (say SU(3)) or increasing the dimensionality of
space-time (say 3+1) are mere technical details and do not pose
conceptual problems. Exactly as happened in the abelian model
\cite{r8,r9,r11}, the nonabelian Thirring model in
the leading $m^{-1}$ approximation gets mapped on to a 
gauge theory, but now it
is nonabelian instead of being
abelian. The physical
sector of this gauge theory, for an arbitrary
Thirring coupling, is abstracted by obtaining both the Gauss
operator that generates the nonabelian gauge transformations and the
physical hamiltonian. The weak coupling limit is then
critically examined. It is shown that in this limit, a covariant
combination of fields in the gauge theory simulating the Thirring
model can be identified with the components of the field tensor
occuring in the Yang-Mills-Chern-Simons (YMCS) theory \cite{r12}
in the limit when the Yang-Mills term vanishes, such that
the physical hamiltonians in the two gauge theories get mapped
on to each other. These results provide a concrete hamiltonian
realisation of the lagrangian approach \cite{r10} where the free 
massive theory was identified with the nonabelian Chern-Simons theory.
Bosonisation identities for the fermionic currents are also
provided.

It is worthwhile to say a few words about our methodology. A
completely different approach than is usually followed for discussing
abelian bosonisation \cite{r5,r6,r7,r8,r9,r11} will be
adopted. In the abelian case master lagrangians \cite{r6,r8,r9,r11,r13}
 are suggested that interpolate
between fermionic and bosonic theories. Such a strategy, however, is
ineffective for obtaining equivalences among nonabelian theories
\cite{r10,r14}. Indeed, as stated earlier, the paper
\cite{r10} that discussed bosonisation of the massive Thirring
model through this technique
yielded explicit results only in the weak coupling limit. Here we
follow a generalised St\"uckelberg like embedding prescription
\cite{r15} for converting second class systems into first class
(gauge) systems. The Thirring model, in the low energy sector, is an
example of a second class system thereby enabling us to exploit the
St\"uckelberg procedure.

The partition function for the SU(2) massive Thirring model is given
by,
\begin{equation}
Z = \int d[\psi, \bar{\psi}] \exp i \int d^{3}x (\bar{\psi}
(i\partial\!\!\!/\, + m) \psi - \frac{g^{2}}{2} j^{a}_{\mu} j^{\mu a})
\label{1}
\end{equation}
where,
\begin{equation}
j^{a}_{\mu} = \bar{\psi}\gamma_{\mu}\sigma^{a}\psi   \label{2}
\end{equation}
is the fermionic current defined by the standard representation of the
Pauli $(\sigma^{a})$ - matrices. As usual, the four-fermion
interaction can be simplified by introducing an auxiliary vector field
$B_{\mu}$,
\begin{equation}
Z = \int d[\psi, \bar{\psi,}B_{\mu}]\exp i \int d^{3}x
(\bar{\psi}(i\partial\!\!\!/\, + m + B\!\!\!\!/\,)\psi +
\frac{1}{2g^{2}} B^{\mu a} B^{a}_{\mu}) \label{3}
\end{equation}
The integration over the fermion fields reduces to the familiar
problem of computing the functional determinant in the presence of an
external field. In general, this expression is nonlocal. Under some
approximation (like the $\frac{1}{m}$ expansion) it yields a local
form. In this low energy regime the leading term is just the nonablian
Chern Simons term so that upto $O(m^{-1})$ \cite{r10},
\begin{equation}
Z \approx \int dB_{\mu} \exp i \int d^{3}x
(\frac{1}{8\pi}\epsilon^{\mu\nu\lambda}
(B^{a}_{\mu}\partial_{\nu}B^{a}_{\lambda} + \frac{\epsilon^{abc}}{3}
B^{a}_{\mu} B^{b}_{\nu} B^{c}_{\lambda}) + \frac{1}{2g^{2}} B^{\mu a}
B^{a}_{\mu}) \label{4}
\end{equation}
Eq.(\ref{4}) represents the partition function\footnote{There is a
technical issue which has been clarified in the discussion below
(\ref{12})} of the nonablian
version of a self dual model considered in \cite{r16,r17,r18,r13}.
 In the abelian case it was possible to exploit the
equivalence \cite{r17,r18,r13} between the self-dual
model and the Maxwell-Chern-Simons theory to obtain a correspondance
between the latter and the massive Thirring model. The above stated
equivalence was demonstrated using either master Lagrangians
\cite{r17,r18,r13,r8} or embedding
presciption \cite{r13}. Since the master Lagrangian approach is
ineffective for nonabelian theories \cite{r14,r10}, we adopt
the embedding approach.

The lagrangian obtainable from (\ref{4}) is,
\begin{equation}
{\cal L} = \frac{1}{8\pi} \epsilon^{\mu\nu\lambda} (B^{a}_{\mu}
\partial_{\nu} B^{a}_{\lambda} + \frac{1}{3} \epsilon^{abc}
B^{a}_{\mu} B^{b}_{\nu} B^{c}_{\lambda}) + \frac{1}{2g^{2}}
B^{a}_{\mu} B^{\mu a} \label{5}
\end{equation}
The equations of motion are given by,
\begin{equation}
\frac{1}{g^{2}} B^{a}_{\mu} + \frac{1}{8\pi}
\epsilon_{\mu\nu\lambda}F^{\nu\lambda, a} = 0 \label{6}
\end{equation}
with,
\begin{equation}
F^{a}_{\mu\nu} = \partial_{\mu} B^{a}_{\nu} - \partial_{\nu}
B^{a}_{\mu} + \epsilon^{abc} B^{b}_{\mu} B^{c}_{\nu} \label{7}
\end{equation}
It is easily seen that the $\mu = 0$ component of (\ref{6}) is an
equation of constraint,
\begin{equation}
\chi^{a} = \frac{1}{g^{2}} B^{a}_{0} + \frac{1}{8\pi} \epsilon_{ij}
F^{ij, a} \approx 0 \label{8}
\end{equation}
where the weak equality is to be interpreted in the sense of Dirac
\cite{r19}. it is clear, therefore, that we are working with a
constrained system. It is simple to extract and classify all the
constraints. The momenta conjugate to $B_{\mu}$ are given by,
\begin{equation}
\pi^{a}_{0} = \frac{\partial {\cal L}}{\partial\dot{B}^{0,a}} \approx
0, \hspace{.5cm} \pi^{a}_{i} = \frac{\partial {\cal
L}}{\partial\dot{B}^{i,a}} = \frac{1}{8\pi}\epsilon_{ij} B^{j,a}
\label{9}
\end{equation}
which are the primary constraints. The second pair of constraints is
just a manifestation of the symplectic structure of the Chern-Simons
term. It is eliminated either by computing the Dirac brackets
\cite{r19} or following the symplectic procedure \cite{r20}, both
of which yield the modified Poisson brackets
\begin{equation}
\{B^{a}_{i}(x), B^{b}_{j}(x')\} = 4\pi \delta^{ab} \epsilon_{ij}
\delta(x-x') \label{10}
\end{equation}
The canonical hamiltonian obtained by a formal Legendre transform of
(\ref{5}) is found to be,
\begin{equation}
H_{c} = \int d^{2}x[\frac{1}{2g^{2}}(B^{a}_{i} B^{a}_{i} + B^{a}_{0}
B^{a}_{0}) - B^{a}_{0}\chi^{a}] \label{11}
\end{equation}
where $\chi^{a}$ is defined in (\ref{8}). Time conserving the primary
constraint $\pi^{a}_{0} \approx 0$ just yields (\ref{8}) as a
secondary constraint. The constraint obtained in the Lagrangian
approach is thereby reproduced in the hamiltonian framework.
Furthermore since,
\begin{equation}
\{\pi^{a}_{0}(x), \chi^{b}(x')\} =
-\frac{1}{g^{2}}\delta^{ab}\delta(x-x')\not\approx 0 \label{12}
\end{equation}
these form a pair of second class constraints.
No more constraints are therefore generated by Dirac's \cite{r19}
iterative prescription. It is not surprising that we have obtained a
second class system since the mass term in (\ref{5}) breaks the gauge
invariance. At this point we remark that the Lagrangian path integral
for (\ref{5}) should contain the constraint (\ref{8}) as a delta
functional in the measure \cite{r21}. However since $B^{a}_{0}$
occurs quadratically in (\ref{5}) the result of integrating out
$B^{a}_{0}$ is the same irrespective of the presence or absence of the
delta functional \cite{r18}. It is for this reason that although the
delta functional does not appear in (\ref{4}), it can still be
regarded as the effective action for the (nonabelian) self dual model
(\ref{5}).

The next step is to convert the second class system (\ref{5}) into a
true gauge (first class) system. A viable approach is to use the
generalised  St\"uckelberg formalism of Kunimasa and Goto \cite{r15}.
The idea is to introduce extra fields $\omega^{a}_{\mu}$ such that
$B^{a}_{\mu} + \omega^{a}_{\mu}$ is gauge covariant. In that case the
embedded lagrangian,
\begin{equation}
\tilde{\cal L} = \frac{1}{8\pi} \epsilon^{\mu\nu\lambda}
(B^{a}_{\mu}\partial_{\nu}B^{a}_{\lambda} + \frac{1}{3}
\epsilon^{abc}B^{a}_{\mu}B^{b}_{\nu}B^{c}_{\lambda}) +
\frac{1}{2g^{2}} (B^{a}_{\mu} + \omega^{a}_{\mu}) (B^{\mu,a} +
\omega^{\mu,a}) \label{13}
\end{equation}
would be gauge invariant. As discussed in \cite{r15} $\omega_{\mu}$
can be constructed from a $2 \times 2$ unitary unimodular matrix $M$
which transforms as
\begin{equation}
M \rightarrow MS \label{14}
\end{equation}
where $S$ is the C-number unitary unimodular matrix that defines the
gauge transformation,
\begin{equation}
B_{\mu} \rightarrow B'_{\mu} = S^{-1}B_{\mu}S + i S^{-1}
\partial_{\mu} S. \label{15}
\end{equation}
Taking $\omega_{\mu}$ to be,
\begin{equation}
\omega_{\mu} = -iM^{-1}\partial_{\mu}M \label{16}
\end{equation}
yields the transformed $\omega_{\mu}$
\begin{equation}
\omega_{\mu} \rightarrow \omega'_{\mu} = S^{-1}\omega_{\mu}S -
iS^{-1}\partial_{\mu}S \label{17}
\end{equation}
Combining (15) and (17) shows that $B_{\mu} + 
\omega_{\mu}$ transforms covariantly,
\begin{equation}
B_{\mu} + \omega_{\mu} \rightarrow B'_{\mu} + \omega'_{\mu} =
S^{-1}(B_{\mu} + \omega_{\mu})S \label{18}
\end{equation}
as required to make (\ref{13}) gauge invariant. A particular form for
$M$ is given \cite{r15} as a function of the Euler angles $\psi,
\theta, \phi$ ;
\begin{equation}
M = \exp (i\psi\sigma^{3})\exp(i\theta\sigma^{1})\exp(i\phi\sigma^{3})
\label{19}
\end{equation}
so that inserting in (\ref{16}) we obtain,
\begin{eqnarray}
\omega^{1}_{\mu} &= &\cos\phi\partial_{\mu}\theta +
\sin\theta\sin\phi\partial_{\mu}\psi \label{20} \\
\omega^{2}_{\mu} &= &\sin\phi\partial_{\mu}\theta -
\sin\theta\cos\phi\partial_{\mu}\psi \label{21} \\
\omega^{3}_{\mu} &= &\partial_{\mu}\phi + \cos\theta\partial_{\mu}\psi
\label{22}
\end{eqnarray}
Note that the Euler angles are to interpreted as field variables. With
(\ref{20}-\ref{22}) defining $\omega^{a}_{\mu}$, the Lagrangian (\ref{13}) is
gauge invariant. This is now explicitly revealed in the hamiltonian
formalism. Apart from (\ref{9}) the other canonical momenta are,
\begin{eqnarray}
\pi_{\theta} &= &\frac{\partial\tilde{\cal L}}{\partial\dot{\theta}} =
\frac{1}{g^{2}}[(B^1_{0} + \omega^1_{0}) \cos\phi + (B^{2}_{0} +
\omega^{2}_{0} \sin\phi] \label{23} \\
\pi_{\psi} &= &\frac{\partial\tilde{\cal
L}}{\partial\dot{\psi}} \nonumber\\
&= &
\frac{1}{g^{2}}[B^1_{0} + \omega^1_{0}) \sin\theta\sin\phi - (B^{2}_{0}
+ \omega^{2}_{0}) \sin\theta\cos\phi \nonumber\\
&+& (B^{3}_{0} + \omega^{3}_{0})
\cos\theta] \label{24} \\
\pi_{\phi} &= &\frac{\partial\tilde{\cal L}}{\partial\dot{\phi}} =
\frac{1}{g^{2}} (B^{3}_{0} + \omega^{3}_{0}) \label{25}
\end{eqnarray}
Substituting the value of $(B^{3}_{0} + \omega^{3}_{0})$ from
(\ref{25}) in (\ref{24}) and then using (\ref{23}) it is possible to
express the covariant combinations in terms of phase space variables,
\begin{eqnarray}
B^{1}_{0} + \omega^{1}_{0} &= &g^{2}[\pi_{\theta}\cos\phi +
\frac{\sin\phi}{\sin\theta} (\pi_{\psi} - \pi_{\phi}\cos\theta)] =
g^{2}L^{1} \label{26} \\
B^{2}_{0} + \omega^{2}_{0} &= &g^{2}[\pi_{\theta}\sin\phi -
\frac{\cos\phi}{\sin\theta}(\pi_{\psi} - \pi_{\phi}\cos\theta)] =
g^{2}L^{2} \label{27} \\
B^{3}_{0} + \omega^{3}_{0} &= &g^{2}\pi_{\phi} = g^{2}L^{3} \label{28} 
\end{eqnarray}
Using the basic Poisson brackets,
\begin{equation}
\{\theta, \pi_{\theta}\} = \{\phi,
\pi_{\phi}\} = \{\psi, \pi_{\psi}\} = \delta(x -
x') \label{29}
\end{equation}
where the arguments of the fields have been suppressed,
it is simple to show that $L^{a}$ defined in (\ref{26} - \ref{28})
satisfy the angular momentum algebra,
\begin{equation}
\{L^{a}(x), L^{b}(x')\} = \epsilon^{abc} L^{c} \delta(x - x').
\label{30}
\end{equation}
The canonical hamiltonian is given by,
\begin{equation}
\tilde{H}_{c} = \int(\pi_{0}B^{0} + \pi_{i}B^{i} +
\pi_{\theta}\dot{\theta} + \pi_{\psi}\dot{\psi} + \pi_{\phi}\dot{\phi}
- \tilde{\cal L}) d^{2}x \label{31}
\end{equation}
Using (\ref{20} - \ref{22}) in conjunction with (\ref{26} - \ref{28})
for eliminating velocities, we obtain,
\begin{equation}
\tilde{H}_{c} = \int d^{2}x[\frac{1}{2g^{2}}(B^{a}_{i} +
\omega^{a}_{i})^{2} + \frac{g^{2}}{2}L^{a}L^{a} - B^{a}_{0}G^{a}]
\label{32}
\end{equation}
where,
\begin{equation}
G^{a} = \frac{1}{8\pi} \epsilon_{ij}F^{a}_{ij} + L^{a} \label{33}
\end{equation}
and satisfies an algebra analogous to (\ref{30}),
\begin{equation}
\{G^{a}(x), G^{b}(x')\} = \epsilon^{abc}G^{c}\delta(x - x') \label{34}
\end{equation}
where use has been made of (\ref{10}). We also exploit the fact that
the new fields and their conjugate momenta have vanishing brackets
with the original phase space variables $B_{\mu}, \pi^{\mu}$. Time
conserving the primary constraint $\pi^{a}_{0} \approx 0(9)$ now yields
the secondary constraint,
\begin{equation}
G^{a} \approx 0 \label{35}
\end{equation}
In contrast to (\ref{12}), $\pi_{0}^{a}$ has vanishing brackets with
$G^{b}$. The involutive algebra satisfied by $G^{a}$ is already given
in (\ref{34}) so that the set of constraints $\pi^{a}_{0}, G^{a}$ is
first class, as expected. The canonical hamiltonian (\ref{32}) is
further simplified on the constraint surface (\ref{35}) to yield the
physical hamiltonian [22],
\begin{equation}
\tilde{H}_{P} = \int d^{2}x[\frac{1}{2g^{2}}(B^{a}_{i} +
\omega^{a}_{i})^{2} + \frac{g^{2}}{64\pi^{2}} F^{a}_{ij} F^{a}_{ij}]
\label{36}
\end{equation}
In the hamiltonian formalism the first class constraint (\ref{35})
plays the role of the Gauss operator and hence is the generator of
time independent gauge transformations. The covariant transformation
law (\ref{18}) is thereby easily verified,
\begin{equation}
\{G^{a}(x), B^{b}_{i} + \omega^{b}_{i}(x')\} =
\epsilon^{abc}(B^{c}_{i} + \omega^{c}_{i}) \delta(x - x') \label{37}
\end{equation}
where we have used (\ref{10}) and,
\begin{equation}
\{L^{a}(x), \omega^{b}_{i}(x')\} = \epsilon^{abc}\omega^{c}_{i}
\delta(x - x') + \delta^{ab}\partial_{i}\delta(x - x') \label{38}
\end{equation}
to compute the basic brackets. The $\mu = 0$ component of (\ref{18})
follows trivially on using (\ref{30}). This completes the conversion
of the second class system (\ref{5}) into first class.

We therefore conclude that the Thirring model bosonises to a gauge
theory whose physical sector is defined by the Gauss constraint
(\ref{35}) and the hamiltonian (\ref{36}). All brackets are canonical
except for (\ref{10}) in which $B_{1}, B_{2}$ are to regarded as the
canonical pair. The bosonisation achieved here is exact in the sense
that the weak coupling limit is not imposed, as was an essential
perquisite in \cite{r10}.

Let us next try to understand the implications of our construction and
see whether it is possible to relate it to some known gauge theory.
The second term in (\ref{36}) resembles the familiar Yang-Mills
contribution while the first involves the new fields. It is clear,
therefore, that (\ref{36}) cannot represent the pure Yang-Mills
theory. In 2+1 dimensions it is possible to think of another familiar
gauge theory - the YMCS theory \cite{r12} whose
dynamics is governed by the lagrangian density,
\begin{equation}
{\cal L} = \frac{g^{2}}{32\pi^{2}} G^{a}_{\mu\nu} G^{\mu\nu,a} -
\frac{1}{8\pi}\epsilon^{\mu\nu\rho}(A^{a}_{\mu}\partial_{\nu}A^{a}_{\rho}
+ \frac{1}{3} \epsilon^{abc}A^{a}_{\mu}A^{b}_{\nu}A^{c}_{\rho})
\label{39}
\end{equation}
where $G^{a}_{\mu\nu}$ is the field tensor (\ref{7}) expressed in
terms of the $A_{\mu}$ field and, for reasons of comparison, we
have taken the same coupling parameter $g$ that appeared in the
Thirring model (\ref{1}). This theory is known to be first class
with the constraints,
\begin{equation}
\pi^{a}_{0} \approx 0, \hspace{1cm} \Lambda^{a} = D_{i}\pi^{i,a} -
\frac{1}{8\pi} \epsilon_{ij}\partial_{i}A^{a}_{j} \approx 0
\label{40}
\end{equation}
where $\pi^{a}_{\mu}$ is the momentum conjugate to $A^{\mu, a}$,
\begin{equation}
\pi^{a}_{i} = - \frac{g^{2}}{16\pi^{2}} G^{a}_{0i} -
\frac{1}{8\pi} \epsilon_{ij}A^{j,a} \label{41}
\end{equation}
The constraint $\Lambda^{a} \approx 0$ is the generator of gauge
transformations enforced by the Lagrange multiplier $A^{a}_{0}$.
This is clearly realised by writing the canonical hamiltonian,
\begin{equation}
H_{c} = \int(\frac{g^{2}}{64\pi^{2}} G^{a}_{ij}G^{a}_{ij} +
\frac{8\pi^{2}}{g^{2}}(\pi^{a}_{i} + \frac{1}{8\pi}
\epsilon_{ij}A^{j,a})^{2} - A^{a}_{0}\wedge^{a})d^{2}x \label{42}
\end{equation}
The physical hamiltonian \cite{r22} is now obtained on the
constraint surface,
\begin{equation}
H_P = \int(\frac{8\pi^{2}}{g^{2}}(\pi^{a}_{i} +
\frac{1}{8\pi} \epsilon_{ij}A^{j,a})^{2} + \frac{g^{2}}{64\pi^{2}}
G^{a}_{ij}G^{a}_{ij})d^{2}x \label{43}
\end{equation}
Comparing (\ref{36}) with (\ref{43}) we see that it is possible
to map the $O(\frac{1}{g^{2}})$ terms using the identification,
\begin{equation}
(B^{a}_{i} + \omega^{a}_{i}) \leftrightarrow 4\pi(\pi^{a}_{i} +
\frac{1}{8\pi} \epsilon_{ij}A^{j,a}) \label{44}
\end{equation}
There are two strong reasons for supporting this identification.
The first is that it is meaningful to compare only gauge
covariant quantities in nonabelian theories since these are related
to
physical variables. Comparing gauge noncovariant objects  - like say
potentials - is meaningless since these are unphysical. The
mapping (\ref{44}) is among gauge covariant quantities with the
R.H.S. just representing the chromoelectric field $G^{a}_{0i}$
(\ref{41}) in the YMCS theory. The second
point is that the mapping is algebraically consistent. Using
(\ref{10}) and (\ref{20} - \ref{22}), it is easy to show,
\begin{equation}
\{B^{a}_{i}(x) + \omega^{a}_{i}(x), B^{b}_{j}(x') +
\omega^{b}_{j}(x')\} = 4\pi\delta^{ab}\epsilon_{ij}\delta(x-x')
\label{45}
\end{equation}
Similarly, using the basic Poisson brackets in the
Yang-Mills-Chern-Simons theory \cite{r12} it can be shown,
\begin{equation}
\{4\pi(\pi^{a}_{i} + \frac{1}{8\pi}\epsilon_{ik}A^{k,a})(x),
4\pi(\pi^{b}_{j} + \frac{1}{8\pi}\epsilon_{jl}A^{l,b})(x')\} =
4\pi\delta^{ab}\epsilon_{ij}\delta(x-x') \label{46}
\end{equation}
which confirms our observation.

Trying to extend the correspondance we find that the $O(g^{2})$
terms in (\ref{36}) and (\ref{43}) do not map on using the
identification (\ref{44}). Thus, in general, the gauge theory to
which the Thirring model bosonises is not the
Yang-Mills-Chern-Simons theory. It is, however, true in the weak
coupling limit $g^{2} \rightarrow 0$ when only the first term in
either (\ref{36}) or (\ref{43}) survives. We therefore conclude
that the SU(2) Thirring model in the weak coupling limit can be
mapped on to the SU(2) Yang-Mills-Chern-Simons theory in the
limit when the Yang-Mills term vanishes. This result was also
obtained earlier \cite{r10} in the master lagrangian approach.

We finally extract the bosonisation identities for the Thirring
current. Following the usual procedure \cite{r6,r8,r9,r10}
 sources $S_{\mu}$ couplied to the current are
introduced in the effective action (\ref{1}),
\begin{equation}
Z = \int d[\psi, \bar{\psi}] \exp i \int
d^{3}x\bar\psi(i\partial\!\!\!\!/\, + m)\psi - \frac{g^{2}}{2}
j^{a}_{\mu} j^{\mu a} + \frac{1}{g^{2}} j^{a}_{\mu} S^{\mu,a})
\label{47}
\end{equation}
where a scaling by $g^{2}$ has been done to make $S_{\mu}$
dimensionless. Then after the introduction of the auxiliary
field,
\begin{equation}
Z = \int d[\psi, \bar{\psi},B_{\mu}]\exp i \int
d^{3}x(\bar{\psi}(i\partial\!\!\!\!/\, + m + B\!\!\!/\, +
\frac{S\!\!\!/\,}{g^{2}})\psi + \frac{1}{2g^{2}} B^{\mu a}
B^{a}_{\mu}) \label{48}
\end{equation}
Replacing $B_{\mu} \rightarrow B_{\mu} - \frac{S_{\mu}}{g^{2}}$ and dropping a
non-propagating contact term yields,
\begin{equation}
Z = \int d[\psi,\bar{\psi},B_{\mu}]\exp i \int
d^{3}x(\bar{\psi}(i\partial\!\!\!\!/\, + m + B\!\!\!/\,)\psi +
\frac{1}{2g^{2}} B^{\mu a}B^{a}_{\mu} - \frac{1}{g^{4}}
B^{a}_{\mu} S^{\mu a}) \label{49}
\end{equation}
Performing the fermionic integration in the large $m$
approximation leads to the nonabelian self-dual model (\ref{4})
so that the Thirring current maps to the basic field
$B^{a}_{\mu}$ in this model. Since the bosonisation identities
are most illuminating in the weak coupling limit $g^{2}
\rightarrow 0$, we henceforth confine our analysis to this
regime. The $B_{\mu}$ field in (\ref{4}) maps to the covariant
combination $(B_{\mu} + \omega_{\mu})$ in the embedded gauge
theory (\ref{13}) so that,
\begin{equation}
j^{a}_{\mu} \leftrightarrow - \frac{1}{g^{2}} (B^{a}_{\mu} +
\omega^{a}_{\mu}) \label{50}
\end{equation}
Using (44) and (41) it is simple to establish the connection of
$j^{a}_{i}$ with the chromoelectric field in the
YMCS theory,
\begin{equation}
j^{a}_{i} \leftrightarrow - \frac{4\pi}{g^{2}}(\pi^{a}_{i} +
\frac{1}{8\pi} \epsilon_{ij}A^{j,a}) = \frac{1}{4\pi}G^{a}_{0i}
\label{51}
\end{equation}
The bosonisation rule for $j^{a}_{0}$ cannot be obtained in this
way since the operator corresponding to $(B^{a}_{0} +
\omega^{a}_{0})$ is not known in the YMCS
theory. We therefore take recourse to an alternative route.
Instead of evaluating (\ref{48}) by translating $B_{\mu}$, the
fermionic integral can be directly computed to yield,
\begin{equation}
Z = \int dB_{\mu}\exp i \int d^{3}x({\cal L} + \frac{1}{8\pi
g^{2}} \epsilon^{\mu\nu\rho} S^{a}_{\mu}F^{a}_{\nu\rho} +
\frac{1}{8\pi g^{4}} \epsilon^{\mu\nu\rho} \epsilon^{abc}
B^{a}_{\mu} S^{b}_{\nu}S^{c}_{\rho}) \label{52}
\end{equation}
where ${\cal L}$ is the usual lagrangian (\ref{5}) for the
self-dual model and $F_{\mu\nu}$ is the field tensor (\ref{7}).
The quadratic term in the sources complicates matters for
extracting a general bosonisation rule. However, if one is
interested in just the correlation functions of $j^{a}_{0}$ then
this quadratic piece drops out and we obtain,
\begin{equation}
j^{a}_{0} \leftrightarrow \frac{1}{8\pi} \epsilon_{ij}
F^{a}_{ij} \label{53}
\end{equation}
Since the basic field $B_{i}$ in the self-dual model maps to
$(B_{i} + \omega_{i})$ in the embedded (gauge) theory which in
turn can be identified with the combination (\ref{44}), the
bosonisation identity (\ref{53}) can be recast in terms of the
fields in the YMCS gauge theory. Equations
(\ref{51}) and (\ref{53}) constitute our bosonisation rules which
are the analogous of the 1+1 dimensional  result of \cite{r23}.

As concluding remarks we mention that a general method for
discussing nonabelian bosonisation (in the low energy regime) of
fermionic theories in higher dimensions has been developed.
Specifically, it was shown that the SU(2) massive Thirring model,
in the leading $m^{-1}$ approximation, was mapped
to a nonabelian gauge theory. The physical sector of this gauge
theory, for an arbitrary
Thirring coupling $g^{2}$,
was explicitly abstracted in the sense that the physical
hamiltonian and the Gauss operator were constructed. The present
approach, therefore, has an advantage over the master lagrangian
approach \cite{r10} where it was problematic  to obtain the
explicit structure of the gauge theory to which the Thirring
model bosonised, unless the weak coupling limit
$g^{2}\rightarrow 0$ was imposed.
Master lagrangians, incidentally, have proved
more fruitful in discussing abelian bosonisation \cite{r6,r8,r9,r11}.

We have next examined the weak coupling limit
 $g^{2}\rightarrow 0$ in details. In this case it was shown
that the corresponding gauge theory
simplified to the Yang-Mills Chern-Simons theory in the limit
when the Yang-Mills term vanishes. A complete mapping was
established by relating a covariant combination of fields in
in the two gauge theories. Finally, bosonisation 
identities for
the fermionic currents were derived.

 It is straightforward to
extend our analysis to gauge groups larger than SU(2).  Then,
instead of just the three Euler variables $\theta, \phi, \psi
(\ref{20} - \ref{22})$, more number of additional fields would
have to be introduced. Similarly, increasing the space-time
dimensionality does not pose difficulties. The evaluation of the
fermion determinant can be carried out in a gauge invariant
fashion. The non gauge invariant piece, therefore, always comes
from the mass term of the auxiliary field. This can be made gauge
invariant by the enlargement prescription discussed here and the
corresponding gauge theory to which the fermionic model bosonises
emerges naturally. As a further future prospect it may be
worthwhile to develop a `smooth nonabelian bosonisation' in
analogy with the 1+1 dimensional case \cite{r3}. Presumably a
more general embedding technique has to be envisaged.

\vspace{2cm}
\centerline{\large\bf Acknowledgement}

One of the authors (RB) would like to thank CNPq (Brazilian
National Research Council) for providing financial support in
carrying out the concluding part of this work. He also thanks
the members of the Instituto de Fisica-UFRJ for their kind 
hospitality.

\end{document}